\documentclass[conference]{IEEEtran}
\IEEEoverridecommandlockouts
\usepackage{cite}
\usepackage{amsmath,amssymb,amsfonts}
\usepackage{algorithmic}
\usepackage{graphicx}
\usepackage{textcomp}
\usepackage{xcolor}
\def\BibTeX{{\rm B\kern-.05em{\sc i\kern-.025em b}\kern-.08em
    T\kern-.1667em\lower.7ex\hbox{E}\kern-.125emX}}
\usepackage{multirow}

\begin{document}

\title{A Study on Broadcast Networks for Music Genre Classification}


\author{
\IEEEauthorblockN{\small Ahmed Heakl$^1$,Abdelrahman Abdelgawad$^2$, Victor Parque$^3$}
\IEEEauthorblockA{\small $^1$\emph{Department of Computer Science, Egypt-Japan University of Science and Technology}, Alexandria, Egypt}
\IEEEauthorblockA{\small $^2$\emph{Department of Mechatronics and Robotics, Egypt-Japan University of Science and Technology}, Alexandria, Egypt}
\IEEEauthorblockA{\small $^3$\emph{Department of Modern Mechanical Engineering, Waseda University}, Tokyo, Japan}
}

%
%
%

\maketitle

\begin{abstract}
Due to the increased demand for music streaming/recommender services and the recent developments of music information retrieval frameworks, Music Genre Classification (MGC) has attracted the community's attention. However, convolutional-based approaches are known to lack the ability to efficiently encode and localize temporal features. In this paper, we study the broadcast-based neural networks aiming to improve the localization and generalizability under a small set of parameters (about 180k) and investigate twelve variants of broadcast networks discussing the effect of block configuration, pooling
method, activation function, normalization mechanism,
label smoothing, channel interdependency, LSTM block
inclusion, and variants of inception schemes. Our computational experiments using relevant datasets such as GTZAN, Extended Ballroom, HOMBURG, and Free Music Archive (FMA) show the state-of-the-art classification accuracies in MGC. Our approach offers insights and the potential to enable compact and generalizable broadcast networks for music classification.
\end{abstract}



\section{Introduction}


Along with the recent developments in Music Information Retrieval (MIR) and the outburst of music streaming services, there has been a rapid increase in the demand for Music Genre Classification (MGC). It is natural for a person's musical taste to favor specific music genres, and the automatic classification of music tracks to a genre is desirable to meet tailored streaming services. Consumers presently access millions of songs via services like Spotify, Tidal, Apple Music and others; all of which needs the autonomous genre classification for the end-user. Also, millions of music files in the internet databases require classification. As such, the automated classification of audio files is expected to tackle the time-consuming and laborious nature of MGC. Machine learning approaches have been used to address the MGC problem since 2002\cite{GTZAN}, achieving the earliest classification accuracy of 60\% and becoming by then the benchmark.

According to the type of input data, MGC is divided into symbolic MGC and direct audio MGC \cite{auto-feature}. The earliest attempts of MGC are of symbolic nature; here, features are manually extracted from the audio signal and then fed to the choice model. As such, after the works of \cite{GTZAN}, researchers began the search for the best representative hand-crafted features for music classification.


Current work in the symbolic MGC field is improving feature extraction to allow the model to identify the genre of a particular track based on the most dominant features representing the track's musical identity. This process requires a lot of knowledge in audio signal processing and music fields. Furthermore, the manually extracted features are not universal, and even if they perform well in a specific task (MGC in this case), they may not perform the same in a different one \cite{CNN+RNN_(parallel)}. The most crucial problem with hand-crafted features is that genres are not clearly defined mathematically, making it hard to determine the best features in the songs representing their genre. Often, most classifiers tend to get confused with similar genres\cite{voting}, e. g., pop and rock, or experimental and instrumental. The attached metadata, which provides information about the samples, will enhance some of these operations. However, as metadata is not always available and because local computing power has dramatically improved, interest in local auto MGC has grown \cite{features_for_music_classification}.

On the other hand, direct audio MGC uses the audio signal as input. It is also possible to perform preprocessing operations on the audio signal before providing it to the model. \cite{spectrograms} and \cite{LBP_textural_features} paved the way for using CNNs in MGC: they were the first to introduce MGC in the visual domain using spectrograms and wavelet packets, respectively, as inputs for the classifier. Having proved effective in image classification\cite{alexnet,inception,inception_v3,resnet,vgg,densenet}, CNNs received the favorable attention of the MGC community due to the ability of auto-extracting features from the audio signal. For CNNs to be used in MGC, the common practice is to use a mel-spectrogram as input, which is much more efficient than raw amplitude signal\cite{Stanford}. Spectrograms are constructed by applying overlapping short windows on the original signal and taking Fast Fourier transform on each window; as such, they are loaded with textural features, being advantageous when used in conjunction with CNNs\cite{hafemann2014forest}.

Although mel-spectrograms are effective mechanisms to encode audio signals, CNNs are generally not good at capturing the temporal features of the audio signal: they are commonly good at extracting spatial features. Even though extracting temporal features is not an issue in image classification, it is a challenging task in MGC\cite{features_for_music_classification}. On the other hand, RNNs can extract temporal features from the spectrogram due to dealing with sequential data. Using both Convolutional and Recurrent Neural Networks enabled the extraction of both spatial and temporal features of the audio signal as in \cite{CNN+RNN_(parallel)} and \cite{CRNN}. \cite{CNNs_best} compared CNNs, LSTMs, and several classical algorithms for MGC: they found that CNNs were the most effective and accurate, outperforming even ensemble classifiers, and being in line with the findings of \cite{evaluation_CNN_spectrogram} and \cite{handcrafted_vs_nonHandcrafted}. Some works like \cite{li2003comparative} suggested that low-level features are important in MGC, \cite{transfer_learning} proved that low-level features are more important than high-level features in MGC as many genres depend heavily on rhythmic patterns and tempo which are basically low-level features. Yet, the relevance of low level features posed a new challenge, since low-level features are usually lost in the depth of CNNs.


The class of Bottom-up Broadcast Neural Networks (BBNN) is the most recent architecture implementing the inception mechanism to extract high-level features while preserving low-level ones from the spectrograms\cite{BBNN}. Although the attractive classification performances were presented in tackling the GTZAN and Ballroom-based datasets, it remains unclear whether the broadcast networks offer competitive generalization abilities on a broader class of music classification settings\cite{BBNN}. Thus, to further explore the performance and generalization landscape of BBNN, we investigate the class of broadcast network architectures and its variants for Music Genre Classification (MGC). In particular, our contributions are as follows:

\begin{itemize}
  \item we propose a broadcast network that improves the localization and generalizability under a smaller set of parameters (about 180k),
  \item we conduct the computational experiments considering well-known and challenging benchmark datasets such as GTZAN, Extended Ballroom, HOMBURG, and Free Music Archive (FMA), showing the state-of-the-art classification accuracies in MGC, and
  \item we investigate twelve variants of broadcast networks discussing the effect of block configuration, pooling method, activation function, normalization mechanism, label smoothing, channel interdependency, LSTM block inclusion, and variants of inception schemes.
\end{itemize}

The above contributions aim at elucidating the performance frontiers of broadcast architectures towards a more generalizable class for MGC. To the best of our knowledge, our studies comparing broadcast networks and their variations in well-known and challenging MGC datasets are the first in the community.




\begin{figure}[t]
    \centering
    \includegraphics[width=0.55\columnwidth]{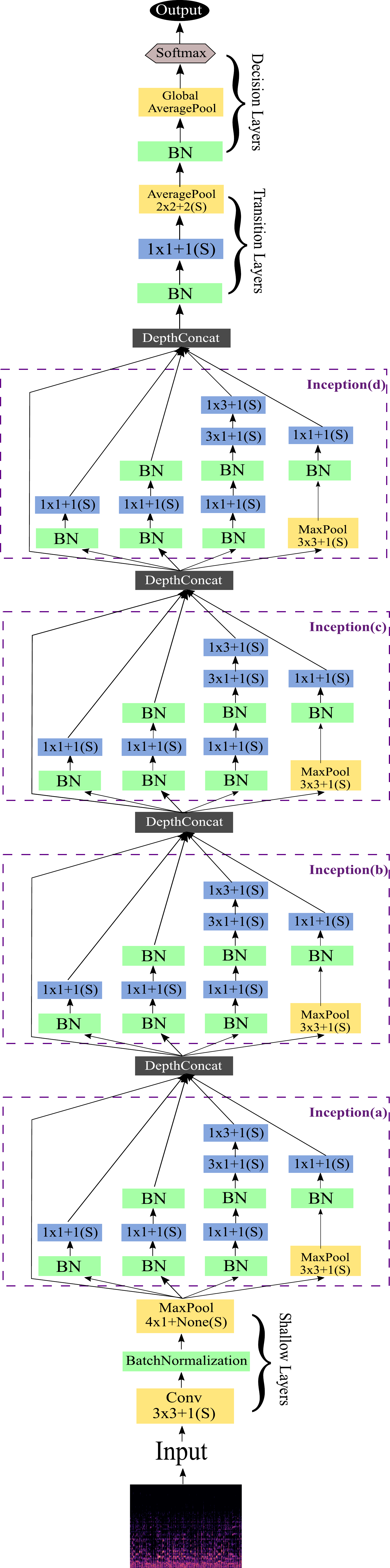}
    \caption{The architecture of our proposed model.}
    \label{fig:Model_Architecture}
\end{figure}

\section{Proposed Architecture}

In this section we present the basic concepts and motivations in our proposed approach.

\subsection{Basic Concept}

Fig. \ref{fig:Model_Architecture} shows the proposed architecture of the broadcast network. Inspired by BBNN\cite{BBNN} and the way in which the ear uses distinct filters to analyze sound characteristics \cite{inception,human_ear}, we propose broadcast networks comprising multiple feature extractors that render different scales of feature maps. And, extending the inception scheme\cite{inception}, we used three convolutions with two different kernel sizes along with a max-pooling layer, and an activation layer for each inception block. Moreover, we used residual connections as they smooth the loss landscape, and allow for deeper architectures\cite{loss_space}. Furthermore, residual connections transmit low-level features to all layers, which are crucial to MGC\cite{transfer_learning}. As such, our module aims at exploiting multiple features in the classifier layer. Compared  to the baseline broadcast-bottom up network\cite{BBNN}, we implement five relevant variations involving: (1) the removal of the 3x3 convolution, (2) the replacement of the 5x5 convolution by a 3x3 convolution, (3) the extension of the number of blocks to four instead of three, (4) the change of the kernel initializer to lecun normal, and (5) the replacement of the asymmetric factorized 3x3 convolutions. And, for each block, we adopt filter sizes of 1x1, 1x1, 3x3, with a 1x1 convolution before the latter to reduce its channel size and create a bottleneck representation \cite{inception}. We also factorized the 3x3 convolutions to 3x1 and 1x3 convolutions as suggested by \cite{inception_v3}. The above-mentioned implementations are motivated by the following notions:

\begin{itemize}
  \item The removal of 3x3 convolutions and the replacement of 5x5 convolutions allowed our model to use more feature maps under the same complexity constraints ($\sim$ 180k parameters), which enriches our model capacity and its ability to capture more textural content from their corresponding spectrograms. Moreover, they provide a better generalization and more depth under the same constraints, which is shown in the training of FMA dataset, Table \ref{table:total_results}.
  \item By adding one additional block, we compensate for the capacity loss from removing the 3x3 convolutions.
  \item Changing the kernel initializer was experimentally proven to increase accuracy significantly.
  \item Finally, the asymmetric factorized convolutions was implemented as inspired by the results from\cite{inception_v3}.
\end{itemize}

\subsection{Network Architecture}\label{AA}

As shown in Fig. \ref{fig:Model_Architecture}, the proposed network consists of $L$ connected modules with residual connections in-between, allowing higher-level layers to receive all feature maps from previous layers. We denote $X_{SL}$ rendered from shallow layers as the input of our module, $L$ is the number of blocks of our module. For simplicity and without loss of generality, we fixed $L = 4$. Thus, the input of the $l$-th block, $l = 1, ..., L$, can be represented as:

\begin{equation}\label{xl}
 X_l = f_1([X_{SL}, X_1, ..., X_{l-1}]),
\end{equation}
where $[X_{SL}, X_1, ..., X_{l-1}]$ refers to the concatenation of the feature maps produced by blocks $0, ..., l - 1$, and $f_1$ is a composite function of all operations in the inception block \cite{BBNN}.

\begin{figure*}[t]
  \centering
  \includegraphics[width=0.98\textwidth]{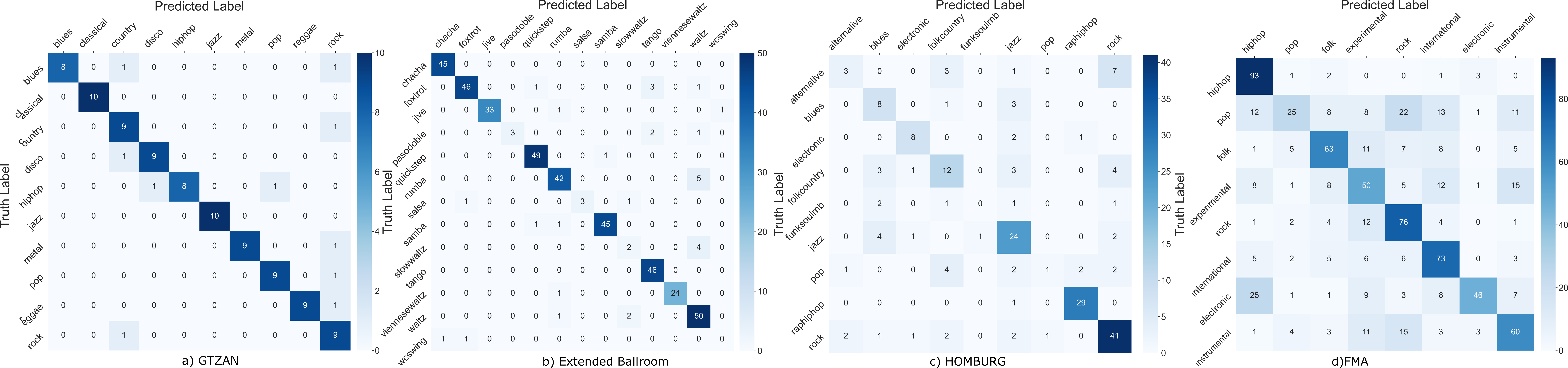}
  \caption{Confusion matrices of our proposed model in all datasets.}
  \label{confusion_matrices}
\end{figure*}

As shown in Fig. \ref{fig:Model_Architecture}, the model comprises 22 convolution layers (5 are 3x3, and 17 are 1x1) with 4 inception blocks to increase the generalizability and allow the extraction of low-level features. As recommended by \cite{batchnorm}, we used Batch Normalization after each convolution layer followed by a ReLU activation; however, we changed the order in the inception inspired by the findings of \cite{identity_mapping}. Batch Normalization does not require regularization and solves the dying ReLU problem. As such, all the layers of the network can be summarized into four different parts: the shallow feature extraction, the proposed module, the transition layers, and the decision layers. The model aims to learn the parameter $\theta$ of a function $F(X_0 | \theta)$ that maps the input spectrogram $X_0$ to its corresponding genre $p$, as follows:

\begin{equation}\label{p}
p = F(X_0 | \theta),
\end{equation}

\begin{equation}\label{p2}
p = f_{DL}(f_{TL}(f_{PM}(f_{SL}(X_0|\theta_{SL}) | \theta_{PM}) | \theta_{TL}) | \theta_{DL}).
\end{equation}

We used a block-based architecture to simplify the design process and have a constant growth rate to facilitate the scaling towards large datasets. Our model has a growth rate of $k_l = k_{l-1} + 4\times f$, where $k_l$ is the output of the $l^{th}$-layer, and $f$ is the number of convolution filters. We fixed the number of convolution filters to be 32. Hence, the output channel dimension of the proposed module is 544, which is reduced in later steps via the transition layer for the classification step.

\begin{table}[t]
    \centering
    \caption{Number of tracks per genre for each Dataset}
    \scalebox{0.75}{
\begin{tabular}{|| cc | cc | cc | cc ||}
\hline
\multicolumn{2}{||l|}{GTZAN}           & \multicolumn{2}{l|}{HOMBURG}   & \multicolumn{2}{l|}{Extended Ballroom} & \multicolumn{2}{l||}{FMA} \\ \hline

\multicolumn{1}{||l|}{Track}   & Genre & \multicolumn{1}{l|}{Track}     & Genre & \multicolumn{1}{l|}{Track}     & Genre  & \multicolumn{1}{l|}{Track} & Genre\\ \hline\hline

\multicolumn{1}{||l|}{Classic} & 100 & \multicolumn{1}{l|}{Electronic} & 113 & \multicolumn{1}{l|}{Cha Cha}        & 455   & \multicolumn{1}{l|}{Rock} & 1000\\

\multicolumn{1}{||l|}{Jazz}    & 100 & \multicolumn{1}{l|}{Jazz}       & 319 & \multicolumn{1}{l|}{Jive}           & 350   & \multicolumn{1}{l|}{International} & 1000\\

\multicolumn{1}{||l|}{Blues}   & 100 & \multicolumn{1}{l|}{Blues}      & 120 & \multicolumn{1}{l|}{Quickstep}      & 497   & \multicolumn{1}{l|}{Folk} & 1000\\

\multicolumn{1}{||l|}{Metal}   & 100 & \multicolumn{1}{l|}{Funk/Soul}  & 47  & \multicolumn{1}{l|}{Rumba}          & 470   & \multicolumn{1}{l|}{Experimental} & 1000\\

\multicolumn{1}{||l|}{Pop}     & 100 & \multicolumn{1}{l|}{Pop}        & 116 & \multicolumn{1}{l|}{Samba}          & 468   & \multicolumn{1}{l|}{Instrumental} & 1000\\

\multicolumn{1}{||l|}{Rock}    & 100 & \multicolumn{1}{l|}{Rock}       & 504 & \multicolumn{1}{l|}{Tango}          & 464   & \multicolumn{1}{l|}{Pop} & 1000\\

\multicolumn{1}{||l|}{Country} & 100 & \multicolumn{1}{l|}{Country}    & 222 & \multicolumn{1}{l|}{Viennese Waltz} & 252   & \multicolumn{1}{l|}{Hip-Hop} & 1000\\

\multicolumn{1}{||l|}{Disco}   & 100 & \multicolumn{1}{l|}{Alternative}& 145 & \multicolumn{1}{l|}{Waltz}         & 529   & \multicolumn{1}{l|}{Electronic} & 1000\\

\multicolumn{1}{||l|}{Hiphop}  & 100 & \multicolumn{1}{l|}{Hiphop}     & 300 & \multicolumn{1}{l|}{Foxtrot}       & 507   & \multicolumn{1}{l|}{}     &    \\

\multicolumn{1}{||l|}{Raggae}  & 100 & \multicolumn{1}{l|}{}           &     & \multicolumn{1}{l|}{Pasodoble}     & 53    & \multicolumn{1}{l|}{}     &    \\

\multicolumn{1}{||l|}{}        &     & \multicolumn{1}{l|}{}           &     & \multicolumn{1}{l|}{Salsa}         & 47    & \multicolumn{1}{l|}{}     &    \\

\multicolumn{1}{||l|}{}        &     & \multicolumn{1}{l|}{}           &     & \multicolumn{1}{l|}{Slow Waltz}    & 65    & \multicolumn{1}{l|}{}     &    \\

\multicolumn{1}{||l|}{}        &     & \multicolumn{1}{l|}{}           &     & \multicolumn{1}{l|}{Weswing}       & 23    & \multicolumn{1}{l|}{}     &    \\ \hline\hline
\multicolumn{1}{||l|}{Total}  & \textbf{1000}  & \multicolumn{1}{l|}{Total}    & \textbf{1886}  & \multicolumn{1}{l|}{Total}    & \textbf{4180}  & \multicolumn{1}{l|}{Total}    & \textbf{8000} \\ \hline
\end{tabular}}
\label{table:Datasets}
\end{table}

\section{Computational Experiments}

We performed computational comparisons using relevant datasets in the community, this section describes our findings.

\subsection{Datasets}

Table \ref{table:Datasets} shows the characteristics of the datasets for MGC used in our study. Basically, we used the well-known and challenging datasets in the community, as follows:

\begin{itemize}
  \item \emph{GTZAN Dataset}. Introduced by \cite{GTZAN}, and being the first publicly available and well-structured benchmark, the GTZAN dataset is a popular dataset in MGC \cite{fma} \cite{Survey_of_Evaluation}. Basically, the GTZAN is a collection of 10 popular genres (blues, classical, country, disco, hip-hop, jazz, metal, pop, reggae, rock) with 100 audio files each, all having a length of 30 seconds.

  \item \emph{Extended Ballroom Dataset}. Assembled for the ISMIR 2004 rhythm description contest, the ballroom dataset includes 698 30-second music clips grouped into eight categories reflecting distinct ballroom dances. The extended ballroom dataset, introduced by \cite{Ballroom}, has 4180 30 second tracks and covers 13 distinct genres, representing an improvement to the original ballroom dataset with about six times the number of tracks with superior audio quality.

  \item \emph{HOMBURG Dataset}. The HOMBURG dataset, presented by \cite{homburg2005benchmark}, is a freely available benchmark dataset for audio classification and clustering. The HOMBURG dataset consists of 1886 songs taken from the Garageband site and classified into nine unbalanced genres. Each sample taken randomly from the song lasts for 10 seconds.


  \item \emph{Free Music Archive Dataset} (FMA). Introduced by \cite{fma}, this dataset is much larger and diverse compared to the datasets mentioned above. Basically, it comprises four subsets: Full, Large, Medium, and Small. FMA is a freely available dataset used to evaluate a variety of MIR tasks. It has 917 GB and 343 days of licensed audio from 106,574 tracks from 16,341 artists and 14,854 albums, organized into 161 genres in a hierarchical taxonomy. Due to hardware constraints in environment, we used the Small set which contains about 8,000 tracks of 30s songs distributed over eight balanced genres (7.2 GiB). As such, compared to the GTZAN and the Extended Ballroom, the above-mentioned selection is comparatively larger, thus we use it to evaluate the ability to generalize and scale.
\end{itemize}




\begin{figure*}[t]
    \centering
    \includegraphics[width=0.98\textwidth]{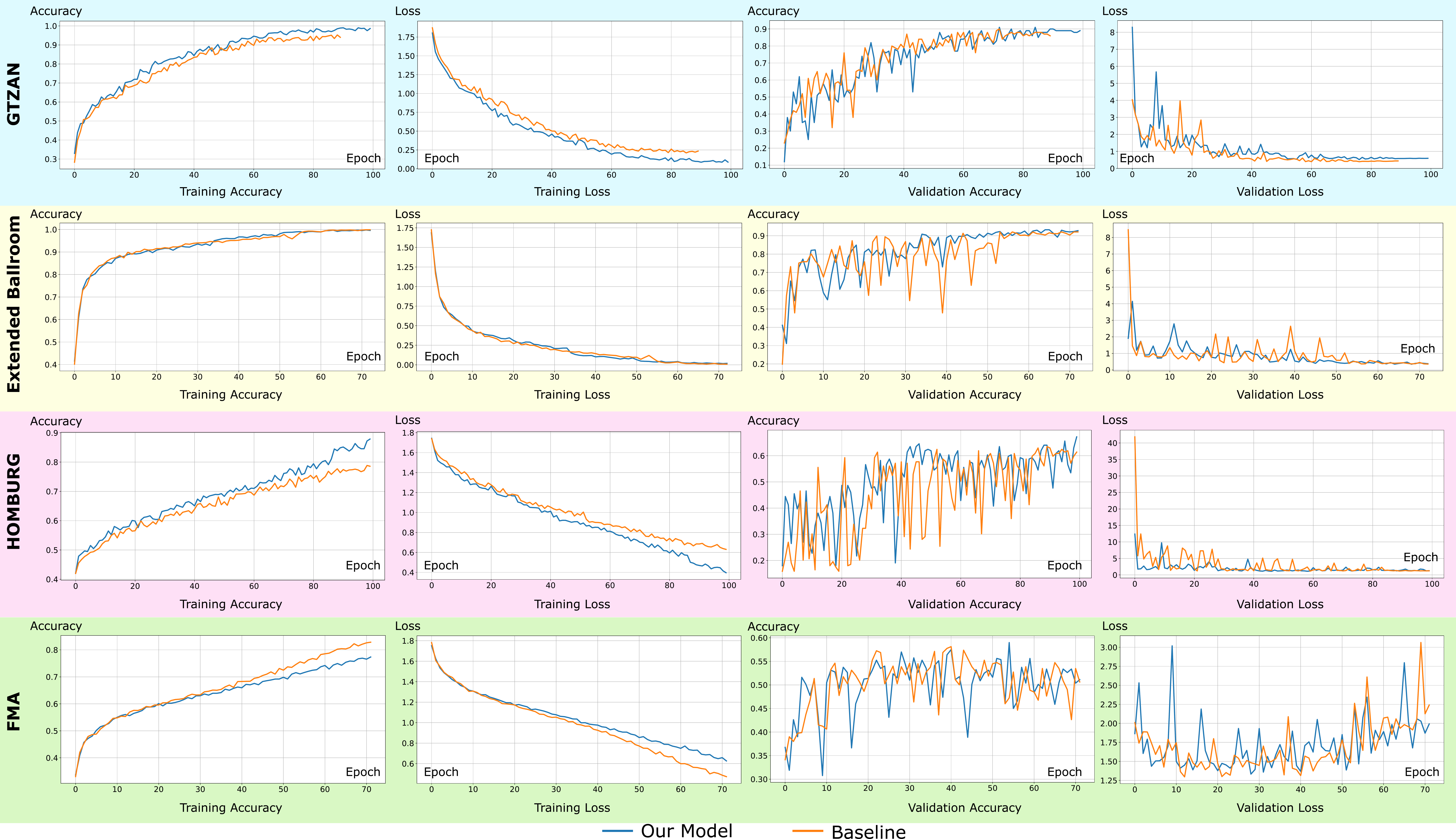}
    \caption{Comparison of the train and validation loss, as well as accuracy between our proposed model and the baseline model for all datasets.}
    \label{fig:curve_gtzan}
\end{figure*}

\subsection{Preprocessing} 

For fairness, we used the same preprocessing technique used by baseline\cite{BBNN}. The Mel-Spectrogram of the audio file was fed to the model as input. Spectrograms were created using a Short Time Fourier Transform (STFT), and then converted into the Mel scale by a logarithmic scale to the frequency axis of the spectrogram in the librosa toolbox \cite{librosa}. 128 Mel-filters (bands) were used to represent the audible spectrum range 0-22050 Hz. We used a frame length of 2048 and a hop size of 1024. Due to the previous setup, the Mel-Spectrogram was of size 646x128.


\subsection{Training Setup}

We used a similar setups for fairness in comparisons with the baseline BBNN model. Our model was built in Keras and trained on Quadro RTX 5000. The model took half an hour to train for 100 epochs on GTZAN. We used a small batch size of 8 since the input resolution is high (647 x 128). Results were measured using a 10-fold validation for each dataset and each variation of the model, and we reported the highest test accuracy. All models were trained using ADAM due to being suitable for noisy gradients, which is the case of high-textural spectrograms\cite{adam}. Inspired by the BBNN model, we set the initial learning rate as 0.01, which was decreased by a factor of 0.5 when the loss stopped improving for three epochs. To avoid overfitting, we set up an early stopping mechanism when the training loss stopped decreasing.

\begin{table}[t]
    \centering
    \caption{Results comparison between our proposed model and baseline}
    \scalebox{0.75}{
\begin{tabular}{|| ll || ll | ll | ll | ll | ll ||}
\hline
\multicolumn{2}{||l}{  } &\multicolumn{2}{|l|}{GTZAN}           & \multicolumn{2}{l|}{Audio Benchmark}   & \multicolumn{2}{l|}{Extended Ballroom} & \multicolumn{2}{l||}{FMA} \\

\multicolumn{2}{||l|}{Model} & \multicolumn{1}{l}{Validation}   & Test & \multicolumn{1}{l}{Validation}     & Test & \multicolumn{1}{l}{Validation}     & \multicolumn{1}{l|}{Test}  & \multicolumn{1}{l}{Validation} & \multicolumn{1}{l||}{Test}\\ \hline

\multicolumn{2}{||l|}{BBNN}
& \multicolumn{1}{l}{90.0} & \multicolumn{1}{l|}{89.0}
& \multicolumn{1}{l}{64.0} & \multicolumn{1}{l|}{61.0}
& \multicolumn{1}{l}{94.7} & \multicolumn{1}{l|}{92.3}
& \multicolumn{1}{l}{58.1} & \multicolumn{1}{l||}{56.9} \\

\multicolumn{2}{||l|}{\textit{\textbf{Ours}}}
& \multicolumn{1}{l}{91.0} & \multicolumn{1}{l|}{\textbf{90.0}}
& \multicolumn{1}{l}{66.1} & \multicolumn{1}{l|}{\textbf{64.0}}
& \multicolumn{1}{l}{93.3} & \multicolumn{1}{l|}{\textbf{93.1}}
& \multicolumn{1}{l}{58.9} & \multicolumn{1}{l||}{\textbf{58.3}} \\

\hline
\end{tabular}}
\label{table:total_results}
\end{table}

\subsection{Results and Discussion}

To show the effectiveness of our proposed model, Fig. \ref{confusion_matrices} shows the confusion matrices of our proposed model in all the datasets, Table \ref{table:total_results} shows the performance comparison (validation-testing) between our proposed model and the baseline, and Fig. \ref{fig:curve_gtzan} shows the learning performance in the training, validation and testing. By observing Fig. \ref{confusion_matrices}, Table \ref{table:total_results} and Fig. \ref{fig:curve_gtzan}, we can observe the following facts:

\begin{itemize}
  \item \emph{GTZAN Results}. As shown by Fig. \ref{fig:curve_gtzan} (a), our model achieved a smaller loss. We also observed that BBNN was unable to differentiate between country, metal, and rock as they were different on the high-frequency components and similar on the low-frequencies. Since our model focuses more on localization, i.e., it allows to capture the high-frequency components, it achieved a much higher accuracy score on the three genres, as shown in Fig. \ref{confusion_matrices} (a). Our model confused hip-hop and pop as the architecture discriminates low-frequency components. Our model boosted the test accuracy by +1\%, as shown  by Table \ref{table:total_results}.

  \item \emph{Extended Ballroom Results}. As the extended ballroom dataset has four times the samples of the GTZAN, as shown in Table \ref{table:Datasets}, the model must use its parameters efficiently. Although both our model and the BBNN show similar training loss values, our model exceeds the performance of the BBNN on the test set but not on the validation set, as shown in Fig. \ref{fig:curve_gtzan} (b). Nonetheless, our model achieves slightly lower validation accuracy than the BBNN model $-1.4\%$, and a higher test accuracy +0.8\%, as shown in Table \ref{table:total_results}. By observing at the confusion matrix in Fig. \ref{confusion_matrices} (b), our model was unable to differentiate betwen Rumba, Slow Waltz, and Waltz. Also, pasodoble was confused with tango, yet we consider this phenomenon due to the low number of samples of the pasodoble genre.

  \item \emph{{HOMBURG Results}}. To evaluate the generalization ability to perform in datasets different than GTZAN and Ballroom Dataset, we experimented with HOMBURG dataset, which has the equivalent size of the datasets used to test the baseline model. Our model outperformed the baseline with +3\% accuracy as shown in table \ref{table:total_results}. We also observed that the model was most successful at identifying jazz, rock, and hip-hop, yet the model failed at identifying the funk/soul genre as it is a hybrid genre. Furthermore, our model confused jazz and blues in some cases as they both have very similar low-frequency components, implying the need for further localization in the convolutional architecture.

  \item \emph{{FMA Results}}. To evaluate the generalizability in a scaled environment, we trained both our proposed model and the baseline in a larger dataset, i.e., the FMA dataset. Since the FMA Dataset has 8000 samples (8 times the GTZAN), both our model and the baseline were unable to capture the variance in the data. However, our model slightly outperformed the baseline with 3\% accuracy, as shown in table \ref{table:total_results}. We argue this phenomenon is due to the increased capacity of our model under the same constraints ($\sim$180k parameters), as our model is one block deeper than the original BBNN model. As shown the learning curves in Fig. \ref{fig:curve_gtzan}, our model has a slightly lower loss in most cases. Our model confused pop with rock which is likely due to the similarity of their low-frequency components, indicating that the model might need further localization to better distinguish the difference. The model also confused the electronic with hip-hop which is also due to the similarity between both genres in almost all features as hip-hop is sometimes categorized as a sub-genre of electronic genre.

\end{itemize}


\subsection{Variants of Broadcast Networks}



We conducted a study to understand further the ability of broadcast networks to outperform state-of-the-art music genre classifiers like those presented in \cite{audeep,nnet2,hybrid_model,transfer_learning,BRNN,CVAF,MFMCNN,Multi-DNN}. In this section, we describe our variations and discuss our findings.


\subsubsection{Removing 3x3 Convolution} 

As shown in Table \ref{tab:variation_accuracy}, by removing the 3x3 convolutions, both the validation and test accuracy in GTZAN dataset dropped down to 84\% and 83\%, respectively. We argue this is due to the decreased model capacity (the number of parameters decreased by 27,744), implying the decreased localization ability without the 3x3 convolutions. As such, the features extracted from the 5x5 convolution became more dominant, which reduced the localization in the features overall.

\subsubsection{Replacing 5x5 Convolution} 
Replacing the 5x5 convolutions with 3x3 convolutions did not affect the model capacity as it gave a validation accuracy of 88\%. However, it causes overfitting as the test accuracy dropped down to 77\%. This is explained by the significant yet uncompensated decrease in the model number of parameters, decreasing the model's ability to generalize.

\subsubsection{Global Average Pooling}
As shown in table \ref{tab:variation_accuracy}, we replaced the global average pooling layer, implemented in the decision layer, with a global max-pooling layer causing the accuracy to decrease drastically. Global average pooling has a few advantages\cite{network_in_network}:
\begin{itemize}
    \item It is more native to the convolution structure by enforcing correspondences between feature maps and categories.
    \item It can be seen as a structural regularizer since there are no parameters to be optimized at this layer and, hence, overfitting is expected to be avoided.
    \item It is more robust to spatial translations of the input as it sums out the spatial information making it useful to be used in the decision layer [25], unlike global max-pooling, which acts as a rectifying unit.
\end{itemize}

\begin{table}[t]
    \centering
    \caption{Variations accuracy on the GTZAN Dataset}
    \resizebox{\linewidth}{!}{
        \begin{tabular}{||l c c||}
             \hline
            Variant & Validation Accuracy & Test Accuracy  \\
             \hline\hline
            Remove 3x3 & 84.0 \%& 83.0 \% \\
            Replace 5x5& 88.0 \%& 77.0 \% \\
            Global Pooling & 76.3 \%& 70.6 \%\\
            Dropout    & 81.5 \%& 78.3 \% \\
            Blocks 1   & 83.8 \%& 82.4 \% \\
            Blocks 2   & 86.0 \%& 82.8 \%\\
            Blocks 5   & 89.0 \%& 86.0 \%\\
            \hline
            SELU Activation         & 89.0 \%& 83.0 \%\\
            Group Normalization     & 78.0 \%& 82.0 \% \\
            Label Smoothing         & 87.3 \%& 84.0 \%\\
            Squeeze \& Excitation   & 85.0 \% & 85.0 \%\\
            LSTM Layer              & 85.0 \%& 82.0 \%\\
            \hline
            Inception Resnet v1     & 87.5 \%& 87.1 \%\\
            Xception                & 89.2 \% & 89.0 \% \\
              \hline
             \hline
        \end{tabular}
    }
    \label{tab:variation_accuracy}
\end{table}

\subsubsection{Dropout} 
Inspired by the inception model\cite{inception}, we added a dropout layer before the fully connected layer mapping from the 32 feature maps to the number of classes. This dropout keeps 60\% of the dense connections. We observed that dropout lowered the validation accuracy to 81.5\%. We argue this phenomenon is due to the small number of parameters of the BBNN model, making it unable to generalize further.


\subsubsection{Number of Blocks} 
As shown in Table \ref{tab:variation_accuracy}, variations of 1, 2, and 5 blocks were evaluated on the BBNN. Although all the variations did not cause overfitting, none of them rendered a higher accuracy than that was obtained from the baseline BBNN with 3 blocks. Decreasing the number of blocks to 1 and 2 decreased the model's capacity making it unable to fit the GTZAN dataset, yet increasing the number of blocks did not increase the model's accuracy.

    \subsubsection{SELU}
    We used the Scaled Exponential Linear Unit (SELU) to avoid the dying RELU problem \cite{selu}, as the SELU activation does not have a zero gradient for input values less than zero. We observed that the model with SELU activation lowered the test accuracy by 3\% as shown by Table \ref{tab:variation_accuracy}. We argue this phenomenon is due to SELU mimicking a soft rectifying unit, implying that piecewise linear functions are better at encoding spectrograms features by stacking frequency bands on top of each other.

    \begin{figure}[t]
        \centering
        \includegraphics[width=\linewidth]{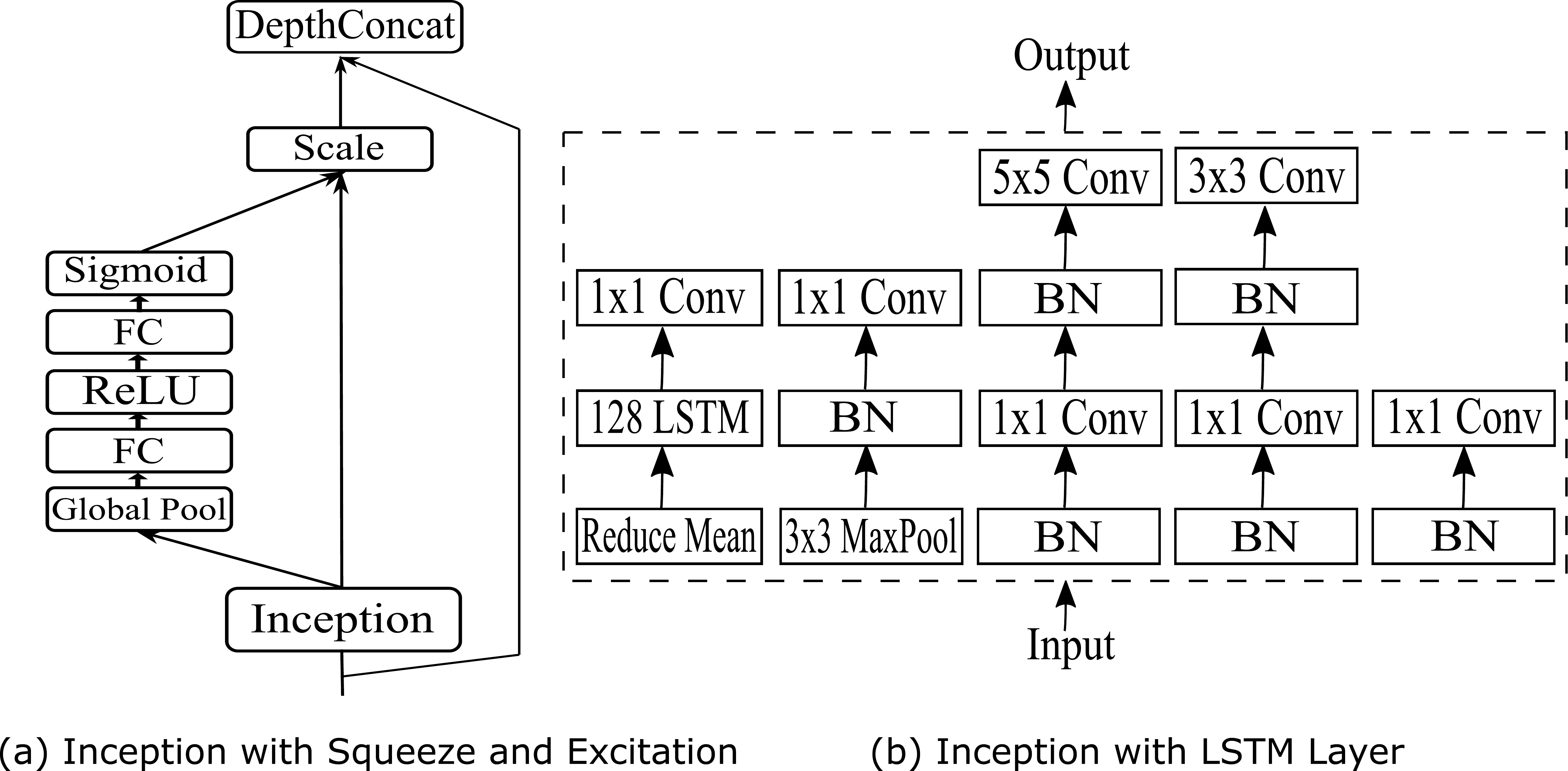}
        \caption{Variants of inception with (a) adding a Squeeze and Excitation Block and (b) adding an LSTM layer.}
        \label{fig:rnn_model}
    \end{figure}

     \subsubsection{Group Normalization}
     Batch normalization adds stochasticity and reduces the model's ability to create discriminative features. Thus, we used Group Normalization (GN) with 16 features per group to keep the performance stable for any batch size, as it calculates the mean and standard deviation over the feature dimension rather than the batch dimension\cite{group_norm}. However, GN reduced the test accuracy by 4\% as shown by Table \ref{tab:variation_accuracy}, implying that batch normalization brings a rather positive effect.


    \subsubsection{Label Smoothing}

   For over-confident models, \cite{inception_v3} proposed a mechanism to regularize the classifier layer by estimating the marginalized effect of label dropout during training. As our training loss was low \(\sim 0.4\), implying the over-confidence of our model, we implemented label smoothing on the input one-hot encoding. We used the same fixed distribution as \cite{inception_v3}: \(u(k) = \epsilon/ K\) with \(\epsilon = 0.1\) and \(K = 10 \) for the GTZAN dataset. We observed that the test accuracy of the model dropped by 2\% as shown by Table \ref{tab:variation_accuracy}, implying that our model is not over-confident, as the low loss suggests.

    \subsubsection{Squeeze and Excitation Block}

    Inspired by \cite{senet}, we modeled the interdependencies between the channels of convolutional features by a Squeeze and Excitation (SE) block after each inception block to improve the efficiency of the feature maps produced by the BBNN model. The inception-SE block is shown in Fig. \ref{fig:rnn_model} (a). We used a reducing factor \(r = 4\) to create a bottleneck representation to reduce the computation power needed. Nonetheless, as shown in table \ref{tab:variation_accuracy}, the test accuracy was reduced slightly by 1\%. We argue this phenomenon occurred due to (1) the increase of the number of parameters with no regularization to tackle overfitting and (2) the lack of depth of our model, implying that feature maps were not necessarily sophisticated enough to emphasize class-specific encodings.

    \subsubsection{Recurrent Neural Networks}
    As recurrent neural networks (RNN) capture temporal features of the inputs, we embedded a Long-Short Term Memory (LSTM) layer into our inception module to act as a temporal feature extractor. The architecture of the inception module with LSTM is shown in Fig. \ref{fig:rnn_model} (b). However, as seen in Table \ref{tab:variation_accuracy}, the accuracy dropped to about 82\%, implying that the added filters from RNN lead to an undesirable increase in model complexity with respect to CNNs for MGC\cite{CNNs_best}.

    \begin{figure}[t]
        \centering
        \includegraphics[width=0.95\columnwidth]{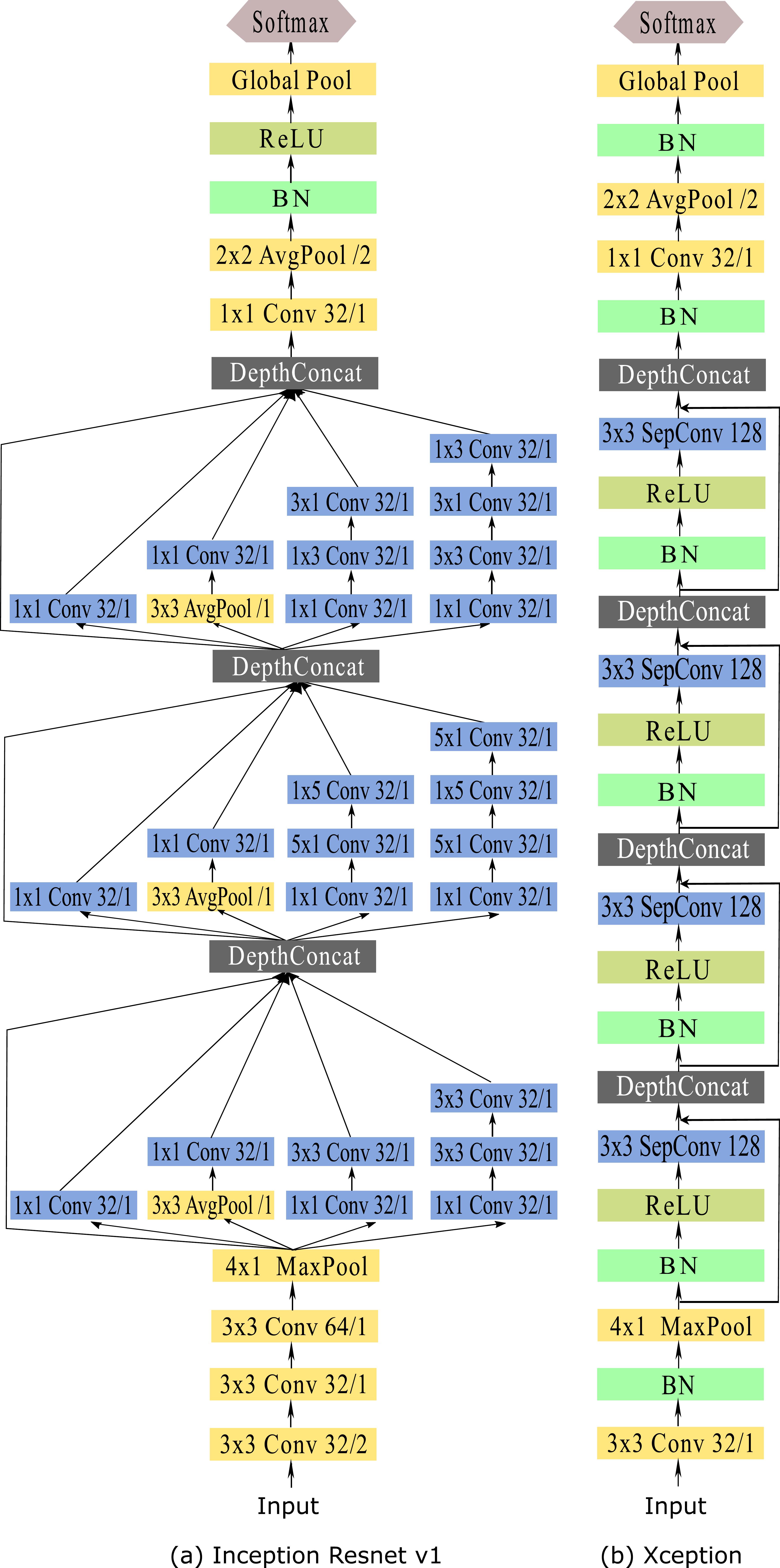}
        \caption{Inception-Based Models: (a) Xception-Based Model (b) Inception-Resnet-Based Model}
        \label{fig:xception_model}
    \end{figure}



    \subsubsection{Inception ResNet v1}

    We constructed the inception-resnet while keeping similar space and computational complexity to the above-described models\cite{inception_v4}. Basically, the model is implemented by replacing each inception block in the BBNN model with its corresponding inception-resnet-v1 block, as shown by Fig. \ref{fig:xception_model} (a). Our inception-resnet model used the same stem block used by the original inception-resnet in \cite{inception_v4}, and the same decision block used by the BBNN. Also, every convolution in each inception-resnet block was preceded with a batch normalization and ReLU activation layers\cite{identity_mapping}. We observed that the constructed inception-resnet reached the convergence at 87.1\% test accuracy, which was 2\% lower than BBNN, as shown in Table \ref{tab:variation_accuracy}. We argue this phenomenon was due to the relatively small size of our model and GTZAN, whereas inception-resnet-v1 targets very large models aided by activation scaling and dimensionality reduction blocks. Thus, contrary to the marketed generalization abilities to different contexts, our results pinpoint that inception-resnet is unable to be easily generalized outside the scope of ImageNet, e.g. to tackle the MGC problem.

    \subsubsection{Xception}
    To avoid using the hand-engineered set of filters in the baseline BBNN model and to better generalize the representation of broadcast convolution filters, we built the xception-based architecture by following the principle of decoupling the mapping of cross-channel correlations and spatial correlations\cite{xception}, whose architecture is shown by Fig. \ref{fig:xception_model} (b). Here, we replaced each inception module with a separable convolution layer to generalize model sparsity\cite{xception}. Our built xception-based model had about 141k parameters, being more compact than the baseline BBNN by 40k parameters, with four separable convolution blocks. Inspired by \cite{identity_mapping}, each block comprises a separable convolution layer preceded by batch normalization and a ReLU activation layer, with residual connections that preserve the low-level features. We observed that the xception-based model achieved a test accuracy of 89\%, by using about 75\% of the number of parameters of the BBNN as shown by Table \ref{tab:variation_accuracy}, showing the effectiveness to achieve comparable performance under smaller space complexity. Further tuning of the configuration of the layers is expected to outperform the baseline\cite{inception_v3}, yet this line of work is out of the scope of this paper.


\section{Conclusions and Future Work}

We have studied the broadcast network architectures for Music Genre Classification (MGC) and proposed a model that improves the localization and generalizability under small number of parameters (about 180k), being potential for practical deployment in embedded devices. Our computational experiments considering well-known and challenging benchmark datasets such as GTZAN, Extended Ballroom, HOMBURG, and Free Music Archive (FMA) has shown the state-of-the-art classification accuracies. Furthermore, we investigated twelve variants of broadcast networks and discussed the effect of block configuration, pooling method, activation function, normalization mechanism, label smoothing, channel interdependency, LSTM block, and inception scheme.

In future work, we will investigate the mechanisms to balance localization and dilation so that the model is able to capture both high and low frequency components. At the same time, we aim at scaling our model to fit large datasets like the FMA dataset. At current, the classification performance in HOMBURG and FMA datasets are still low, suggesting that broadcast networks still lack the desirable generalization abilities. Furthermore, we plan to explore different representations of audio files; although spectrograms are convenient and straightforward to use in MGC, there is potential to explore other representations such as scalograms and wavelet scattering transforms.



\bibliographystyle{IEEEtran}
\bibliography{Reseach_References}

\end{document}